# Minimizing the programming power of phase change memory by using graphene nanoribbon edge-contact

Xiujun Wang[1,2,3†], Sannian Song[1,2†], Haomin Wang[1,2,3*], Tianqi Guo[1], Yuan Xue[1,2], Ruobing Wang[1,2], HuiShan Wang[1,2,3], Lingxiu Chen[1,3], Chengxin Jiang[1,3,4], Chen Chen[1,2,3], Zhiyuan Shi[1,3], Tianru Wu[1,2,3], Wenxiong Song[1,2], Sifan Zhang[1], Kenji Watanabe[5], Takashi Taniguchi[6]，Zhitang Song[1,2*], Xiaoming Xie[1,2,3,4]

[1] State Key Laboratory of Functional Materials for Informatics, Shanghai Institute of Microsystem and Information Technology, Chinese Academy of Sciences, 865 Changning Road, Shanghai 200050, P. R. China.

[2] Center of Materials Science and Optoelectronics Engineering, University of Chinese Academy of Sciences, Beijing 100049, P. R. China.

[3] CAS Center for Excellence in Superconducting Electronics (CENSE) Shanghai 200050, P. R. China.

[4] School of Physical Science and Technology, ShanghaiTech University, Shanghai 201210, P. R. China.

[5] Research Center for Functional Materials, National Institute for Materials Science, 1-1 Namiki, Tsukuba 305-0044, Japan.

[6] International Center for Materials Nanoarchitectonics, National Institute for Materials Science, 1-1 Namiki, Tsukuba 305-0044, Japan.

† These authors contribute equally to this work.

* Electronic mail: hmwang@mail.sim.ac.cn, ztsong@mail.sim.ac.cn

**Abstract:** Nonvolatile phase change random access memory (PCRAM) is regarded as one of promising candidates for emerging mass storage in the era of Big Data. However, relatively high programming energy hurdles the further reduction of power consumption in PCRAM. Utilizing narrow edge-contact of graphene can effectively reduce the active volume of phase change material in each cell, and therefore realize low-power operation. Here, we demonstrate that a write energy can be reduced to about ~53.7 fJ in a cell with ~3 nm-wide graphene nanoribbon (GNR) as edge-contact, whose cross-sectional area is only ~1 nm$^2$. It is found that the cycle endurance exhibits an obvious dependence on the bias polarity in the cell with structure asymmetry. If a positive bias was applied to graphene electrode, the endurance can be extended at least one order longer than the case with reversal of polarity. The work represents a great technological advance for the low power PCRAM and could benefit for in-memory computing in future

**Main text**

Phase-change random access memory (PCRAM) is a promising circuit-building block for non-von Neumann computing architectures by combining storage and computing functions. It is equipped with desirable properties for nonvolatile, high device density, high programming speeds, and long switching lifetime[1,2]. These properties are the key enablers of components for in-memory computing in highly data-centric applications. Therefore, in terms of materials and device structure, new phase change devices are highly desired to enable high throughput, area efficient and energy efficient information processing. Both capitalizing on new phase change materials (PCMs)[3] and applying an incubation field[4] can lead to higher operating speed, while stacking phase-change multilayered hetero-structure could extend switching lifetime[5]. The greatest challenge faced by PCRAM at this moment is how to reduce its power consumption further. The scaling on phase-change cells not only can enable the integration of more memory devices, but also reduce the programming energy.

The conventional memory cell exhibits a relatively high power consumption due to its large contact area between PCM and heating electrode[6]. The cutting-edge technologies in semiconductor industry can scale memory cells only down to sub-7-nm, but the contact area is still in tens of nm$^2$. Creating a gap in carbon nanotube can decrease the active volume of phase change materials by orders and then reduce the power consumption[7-10]. Nevertheless, the technique needs extensive skills and complex processes in both nano-gap fabrication and deposition of phase change material. Using blade electrodes is another approach to reduce the materials needed to be heated up, and then consume less energy. However, the cross-sectional area of the memory cells with the most advanced blade electrode still exceeds 40 nm$^2$ [11,12].

Graphene is a two-dimensional semi-metal with a thickness of a single atom (~0.335 nm)[13,14]. It can make one dimensional electrical contact to other materials by edge-contact[15,16]. In addition, graphene is chemically inert and thermally stable, with a current-carrying capacity of more than $10^9$ amps per square centimeter[17,18]. The merits are highly desired in electrodes for phase change programming. Therefore, graphene can serve as the thinnest blade-electrode for PCRAM cells in nature. Recently, lateral PCRAM cells with patterned GNR electrodes were demonstrated in Stanford University[19]. However, the power consumption of the cells is still one order of magnitude greater than those with CNT electrodes due to the challenges in process control over the quality of both GNR-GST interface and GST material. Encapsulation of GNRs between *h*-BN flakes can greatly preserve their intrinsic property by keeping them from the ambient environment during programing operation. Besides, the *h*-BN flakes in a suitable thickness can serve as a tiny "heat sink" in programing process because of its high in-plane thermal conductivity[20,21]. Bringing the edges of both GNR and *h*-BN heat sink to contact phase change material builds up a new structure for PCRAM cells.

In this report, we are using edges of graphene nanostructure as blade-contact to memory cells. The dimensions of the cells are in general defined by the width of edge contact of graphene (~3 nm to 2 μm). As the edge is the location with the highest E-field and Joule heating in graphene, switching of PCMs occurs only near the edge contacts. It is found that the switching speed reaches 5 ns in memory cell with graphene edge-contact. Its operation duration is close to $6\times10^6$ if a proper bias polarity was applied. The write energy of the memory cell decreases to ~53.7 fJ when the contact width decreases to ~3 nm. The corresponding cross-sectional area of the edge-contact is only ~1 nm$^2$. Its operation duration is close to $3\times10^5$, which allows reliable iterative programming operations. In addition, a D flip-flop prototype working under 2.5 MHz was demonstrated in a memory cell with graphene nanoribbon (GNR) edge-contact. Here, Ge$_2$Sb$_2$Te$_5$ (GST) is selected as PCM. The detailed process for fabrication of a memory cell with edge-contact is described in methods and supplementary information (Fig. S1).

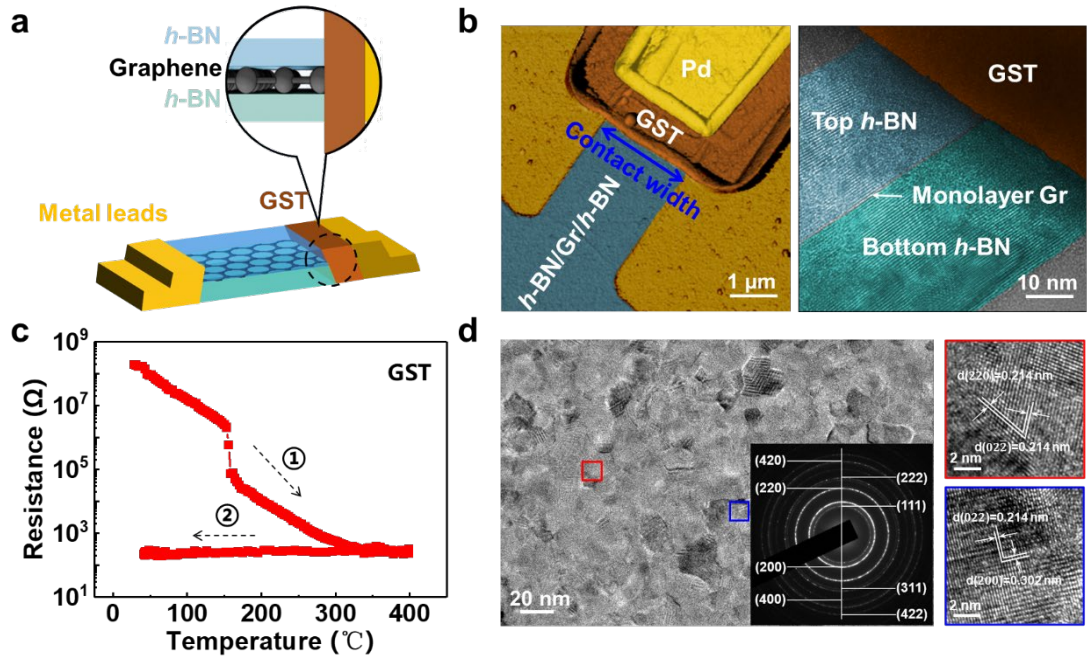

**Fig. 1 Memory cell with graphene edge-contact.** (**a**) Schematic of the cell. (**b**) AFM image of a memory cell where palladium (Pd), GST and *h*-BN/Gr/*h*-BN are painted with gold, brown and cerulean, respectively. The scale bar is 1 μm. The HRTEM image on the right shows a cross-sectional view of the edge-contact, the scale bar is 10 nm. (**c**) The resistance variation of a 60 nm-thick GST film in a cycle of annealing. The heating and cooling rates are ~20 °C/min and ~100 °C/min, respectively. (**d**) TEM investigation on another GST film annealed at 260 °C, and the insert shows the corresponding SAED pattern. Zoom-in view on two specific crystal grains framed are given on the right.

Fig. 1a shows a schematic of a memory cell by using graphene edge as one-dimensional contact. Its AFM image is given on the left of Fig. 1b. The width of edge-contact is about 2 μm. A high-resolution transmission electron microscopy (HRTEM) image shows a cross-sectional view of edge-contact on the right of Fig. 1b. The blue and green areas represent the capping and bottom layer of *h*-BN, respectively. The grey contrast comes from their different lattice orientations. Position of the graphene layer is also pointed out in Fig. 1b. Fig. 1c shows the resistance-temperature relationship of GST, which exhibits a crystallization transition at the temperature of ~155.7 °C. The data retention of the RESET state in GST cells is expected to be ~80.2 °C for 10 years by Arrhenius fitting (Fig. S2). As shown in Fig.1d, the selected area electron diffraction (SAED) pattern of the GST film exhibits a face centered cubic (FCC) structure. The structure was further confirmed with HRTEM images of nanocrystal grains.

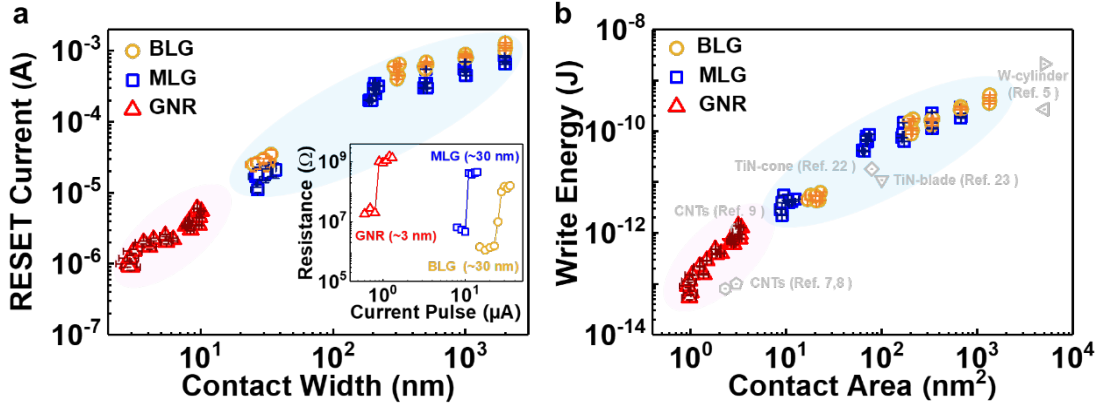

**Fig. 2 Scaling trend of power consumption.** (**a**) RESET current *versus* width of edge contact in memory cells. Insert shows RESET process driven by current pulses in the memory cells with different edge-contact. (**b**) Write energy as a function of contact area. The yellow circle, blue square and red triangle represents the results from the 74 cells with edge contact of BLG (26 pieces), MLG (29 pieces) and GNR (19 pieces), respectively. The other symbols in gray are adapted from literatures[5,7-9,22,23].

Bilayer graphene (BLG), monolayer graphene (MLG), and GNR in different widths are used as the edge-contact of memory cells. The narrowest pattern of graphene defined by lithography is in width of ~30 nm. To narrow the contact further, GNRs embedded in *h*-BN[24] was used as edge-contact to PCMs. The width of GNRs used here is less than 10 nm. As the SET energy $E_{SET}$ is almost three orders lower than the RESET energy $E_{RESET}$, it is always ignored when calculating the power consumption of memory cells[25]. The memory cells were firstly annealed at 260 °C in order to set to low resistance state (LRS, sometimes called as ON state). Subsequently, a series of current pulses with a fixed duration (100 ns) were sourced to the cell *via* graphene edge-contact with the magnitude of the pulses gradually increasing. Simultaneously, the resistances of memory cells were recorded. And the results are shown in the insert of Fig. 2a. The action to set the cell from LRS to high resistance state (HRS, sometimes called as OFF state) is marked as RESET, this switching is consistent with quick melting and quenching of GST, turning the material to amorphous phase. Here, the minimum of current pulse switching the cell to HRS is defined as RESET current. (More results about RESET current are shown in Fig. S3) The variation of RESET current with respect to the width of edge-contact are plotted in Fig. 2a. It is found that RESET current monotonously decreases with the shrink of contact width, while the cell resistance measured increases. It is also found that the RESET current of memory cells with BLG contact is almost twice of that of cells with MLG contact in the same width. The RESET current measured is ~11 μA in the cell with ~30 nm-wide edge-contact of MLG. And the RESET current in a pulse of 100 ns obtained is ~0.9 μA in a memory cell with a ~3 nm-wide GNR edge-contact.

Subsequently, the contact-area dependence of power consumption ( $E = \int_0^{\Delta t} IV\, dt$ )

of each memory cell is extracted, here *I* represents RESET current, *U* is RESET voltage, and Δ*t* is a time symbol for pulse duration. Transient RESET voltage measured is recorded once the cell completes the RESET action. The relationship of write energy *versus* contact area in memory cells is plotted in Fig. 2b. The power consumption of low power PCRAMs reported in literatures were also included for comparison. It is found that the write energy decreases with the contact area. The power consumption decreases to ~53.7 fJ for the cell with ~3 nm-wide GNR edge-contact. The value is obviously smaller than those reported in literatures, enabled by the tiny amount of GST addressed with the GNR electrode. Decreases of power consumption attribute to the shrink of contact area between electrodes and GST.

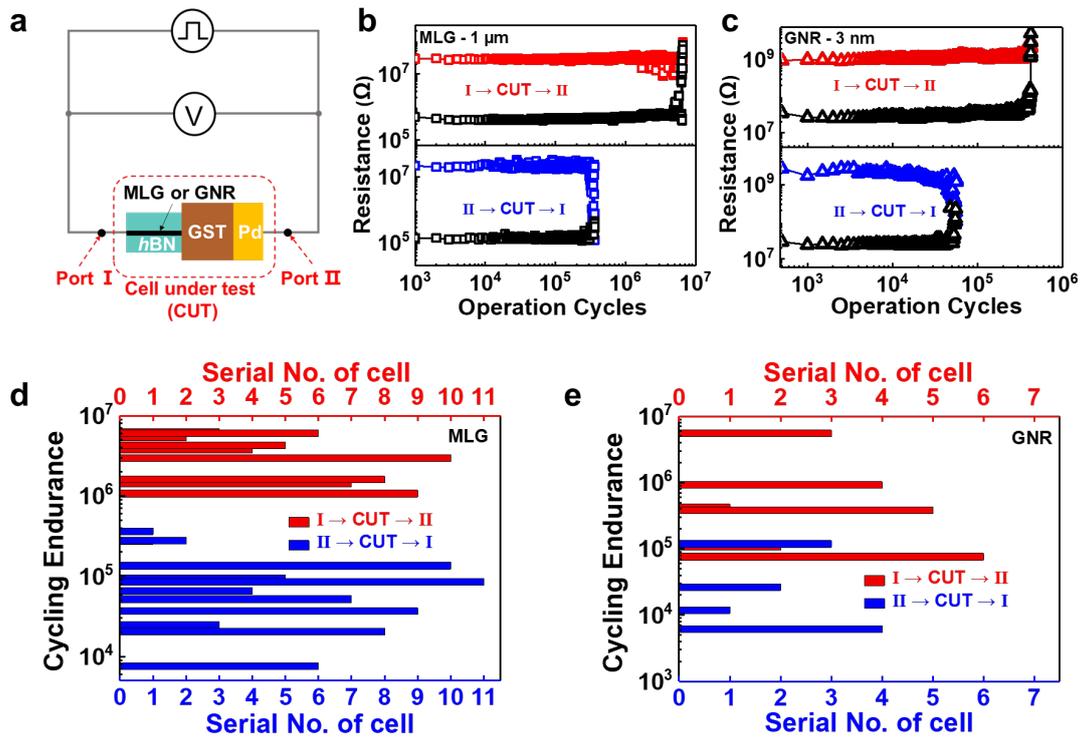

**Fig. 3 Endurance dependence on voltage polarity.** (**a**) Schematic of the cell and measurement layout. (**b**) Endurance test in cells with MLG edge-contact. The data shown in the upper diagram were obtained with 1.5 V/100 ns SET pulse and 2.5 V/100 ns RESET pulse in a cell (*#MLG 74*) with ~1 μm wide MLG edge-contact, while those shown in the bottom diagram were token with 1.5 V/100 ns SET pulse and 2.8 V/100 ns RESET pulse in another cell (*#MLG 55*) in similar configuration. (**c**) Endurance investigation of cells with ~3 nm wide GNR edge-contact. The results in the upper diagram were obtained with 0.5 V/100 ns SET pulse and 1 V/100 ns RESET pulse (*#GNR 103*), while those shown in bottom diagram were token with 0.6 V/100 ns SET pulse and 1.2 V/100 ns RESET pulse in another cell (*#GNR 108*) in similar configuration. Similarly, voltage pulse from Port I→CUT→II leads to long endurance and "Stuck RESET" failure as shown in the cell. Applying a voltage pulse from Port II

to I through CUT results in short endurance and failure in a low resistance state ("Stuck SET") as shown in the Cell. **(d, e)** Statistics on the cycle endurance for MLG and GNR edge-contact memory cells, respectively.

The cells using MLG/GNR as edge-contact are physically asymmetric in their device structure. The physical asymmetry could have a significant influence on the device endurance under different voltage polarity. Cycle-endurance is extensively investigated in the cells with edge contact. The voltage for read-operation always is kept at 0.1 V unless otherwise noted. Schematic of the cell and measurement setup is shown in Fig. 3a. The results about cycle-endurance measured in ~1 μm MLG edge-contact cells and GNR edge-contact cells in ~3 nm width are plotted in Fig. 3b and 3c, respectively. The memory cell with ~1 μm MLG edge-contact (*#MLG 74*) exhibits an endurance close to $6\times 10^6$ (Fig. 3b), and the ~ 3 nm GNR edge-contact memory cell (*#MLG 103*) shows an endurance up to $3\times 10^5$ when the voltage pulse is applied from Port I to II (Fig. 3c).

Meanwhile, when a pulse voltage was applied from Port II to I through CUT, the endurances become just around $10^5$, as shown in Fig. 3b and 3c. It is also noted that the voltage polarity has a strong effect on failure modes. When voltage pulse was applied from Port I→CUT→II, the cell fails in a high resistance state ("Stuck RESET"). Reversal of voltage polarity always leads to the failure of "Stuck SET".

A systematic statistical investigation on cycle endurance for MLG and GNR edge-contact cells is carried out, and the results are shown in Fig.3d and 3e, respectively. As shown in Fig.3d and 3e, a relatively long endurance can be achieved in most of the cells with edge contact if the voltage pulse was applied from Port I to II through CUT. Cells applied with voltage pulse direction (Port I→CUT→II) are all found to fail in "Stuck RESET", while those in reversal of voltage polarity fail in "Stuck SET". (See Fig. S5 in supplementary information) Fig. S6 shows a cross-sectional TEM image of the cell after "Stuck RESET" failure. The voids (profiled by white dash line Fig. S6) are found at the contacts between graphene-edge and GST, it is generally recognized as the cause for "Stuck RESET".

A cross-sectional TEM image and elemental EDS maps of the cell with "Stuck SET" failure after endurance test was shown in Fig. S7. As shown in Fig. S7f, 7g and 7h, Ge, Sb and Te element extended into the interface of *h*-BN/graphene by emigration. As shown in Fig. S7i, the counts of Ge elements are slightly more than Sb and Te in the active region (colored with red). The composition of GST in the gray region of Fig. S7i is similar to the composition of GST (Fig. S7j) in the inactive region marked in Fig. S7a. This indicates that the size of the active region is ~10 nm. When the voltage pulse was applied from Port II→CUT→I, elements in GST emigrate into the interlayers of graphene/*h*-BN under the electric and thermal fields, and then form a small confined active region. In this tiny region, the element segregation of the GST under cycle test greatly degrades the cycle endurance with "Stuck SET" failure.

The resistance drift was also measured in the cells with graphene edge-contact (Fig. S8), and the coefficient of resistance drift is found to be close to 0.02, which is significantly smaller than that in memory cells with cylinder electrode ($v = \sim 0.1$)[26]. We attribute the small resistance drift to the extremely high in-plane thermal conductivity of graphene[27] and $h$-BN flakes[20,21], contacting the active region of phase change. The graphene and $h$-BN flakes help to greatly reduce the thermal relaxation of GST in the programing process, and thus suppresses the resistance drift.

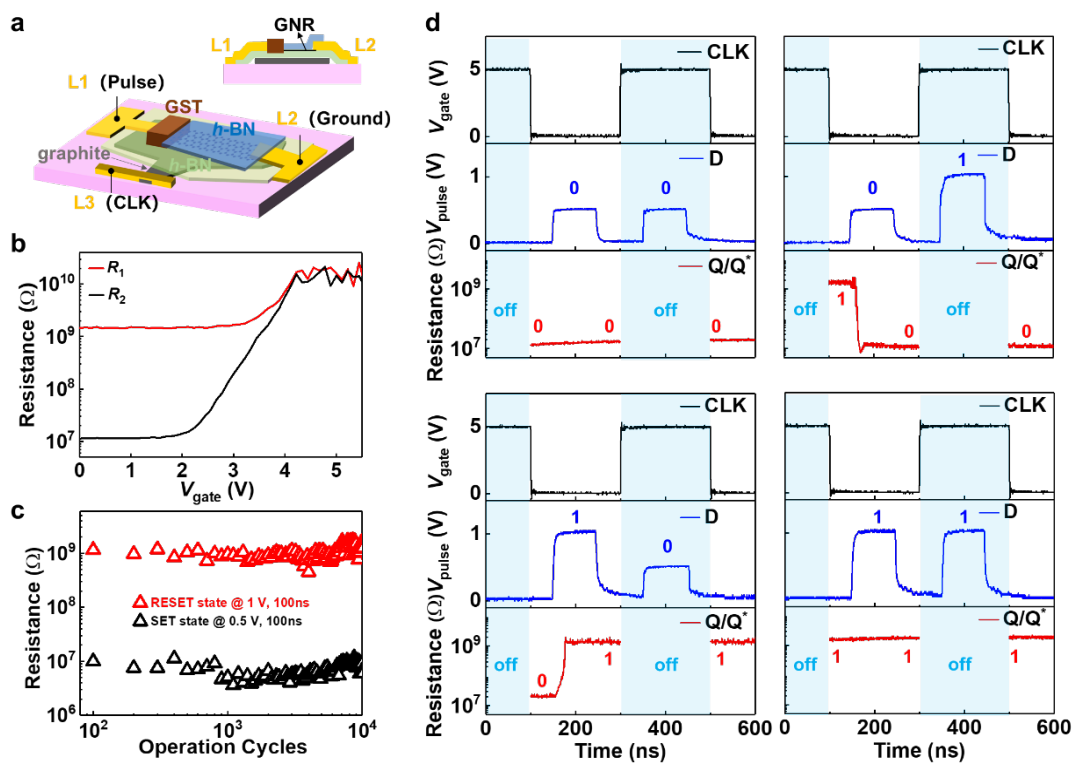

**Fig. 4 A prototype of D flip-flop made from a memory cell with GNR edge-contact.** (a) Schematics of the prototype and the measurement setup. (b) The cell resistance *versus* gate voltage ($V_{gate}$) at 300 K. (c) Cycling endurance of the memory cell. Both SET (0.5 V)/RESET (1 V) signals are in a pulse width of 100 ns. (d) Demonstration of logic functions in the D flip-flop.

As the nonvolatile memory cells can perform the function of resistive switching, we demonstrated the logic function of D latch based on the memory cells with edge-contact. The results are shown in Fig. S9. Fig. S9a displays the logic function of D latch. Input signals can modify the magnitude of output resistance. It is always expressed as the switching of logic states in the cell. Here, "D", "Q", and "Q*" denote input, initial state, and subsequent state, respectively. Both pulse of high voltage (2.5 V) and HRS are defined as logical "1", while pulse of low voltage (1.5 V) and LRS are defined as logical "0". As shown in Fig. S9a, if no voltage pulse is applied (D = "zero input"), the subsequent state (Q*) remains the initial state (Q). When a voltage pulse in 1.5 V (D = logical "0") is sourced, the cell switches from the HRS to LRS. Similarly, a voltage pulse of higher amplitude (2.5 V) (D = logical "1") will switch the cell from the LRS to HRS. The voltage for reading the cell resistance always remains less than 0.1 V unless otherwise noted. The corresponding truth table and logic diagram are shown in Fig. S9b and S9c, respectively.

On this basis, we prepared a D flip-flop made from a memory cell with GNR edge-contact. The fabrication process for D flip-flop based on the memory cell with edge-contact of GNR is shown in Fig.S10. Fig. 4a shows a schematic and measurement setup for the D flip-flop. During operation, L1(connected to GST) and L3 (connected to the graphite which serves as back-gate) are selected as the input (D) and clock signal (CLK), whereas L2 (connected to GNR) is grounded. To investigate the gate tunability of the cell with GNR edge-contact, a small lateral voltage (0.1 V) was applied through L1 and L2 to measure the resistance by sweeping $V_{gate}$ from 0 V to 5.5 V. Fig. 4b shows the resistance of the cell *versus* $V_{gate}$ at 300 K. The red and black line were measured when the memory cell in HRS and LRS, respectively. It is found that the channel of the cell with GNR electrode is in ON-state when $V_{gate}$= 0 V and OFF-state when $V_{gate}$= 5 V (see the $R$-$V_{gate}$ curve in Fig. S11a, the GNR transistor exhibits $R_H$ ($V_{gate}$=0 V)/$R_L$ ($V_{gate}$=5 V) ~$10^5$ at 300 K).

In order to test its cycling endurance, we set 0.5 V/1 V as SET/RESET voltage with a pulse width of 100 ns to complete the reversible switch from HRS (~$10^9$ Ω) to LRS (~$10^7$ Ω), as shown in Fig. 4c. The memory cell shows a cycling endurance more than $10^4$.

Fig. 4d shows a dynamic logic function of the D flip-flop under a clock (CLK) signal of 2.5 MHz. The test conditions are the same as that in Fig. 4c. The input D in a voltage pulse of 0.5 V (1V) is regarded as logical "0" ("1"). Q and Q* represent the initial and subsequent resistance state of the memory cell, respectively. The LRS of cell means logical "0" while HRS in cell is considered as logical "1". When CLK is in 0 V (logical "1"), Q* becomes logical "0" or "1" regardless of the initial resistance state, by following D. However, when CLK is in 5 V, Q* will not change but remains in its initial state of Q. The transient change of resistance can be measured directly by an oscilloscope, the details about measurement is shown in Fig. S12. It is noted that the switch between two states takes ~20 ns, greatly shorter than 100 ns of input pulse. This indicates that the excrescent electrical energy converts to the Joule heat without causing amorphous–crystalline transitions. The corresponding truth table is shown in Fig. S13b. D flip-flop normally takes lots of transistors to implement in the traditional CMOS technology. Here, the memory cell with GNR edge-contact realized the logic function of a D flip-flop, which just is composed of PCMs and GNR channel. In addition, the D flipflop is of the capability of nonvolatile memory. It could be used to produce low power arithmetic/logic units in nonvolatile random-access memories.

Scaling PCRAM cells to low dimension could produce more new physics to boost their performance if the GST was engineered or changed to other PCMs. A cell with edge-contact achieves a SET speed of ~6 ns at ~0.45 V, and it exhibits an ultra-low power consumption of ~53.7 fJ in a high endurance of $3\times10^5$ when the width of contact approaches 3 nm. The power consumption of each cell is almost several orders of magnitude lower than that in the state of art PCRAM cells[28-30], and even half lower than that in the cell addressed by a CNT gap[8]. The edge-contact geometry in PCRAM devices could help explore the dynamic switching process of phase-change material at the atomic level and engineer different materials, and then improve their electrical performance. From an application point of view, the performance of PCRAM cells made from graphene edge-contact allows them serving as nonvolatile DRAM[31] in the future if their endurance is improved further via engineering PCMs or device configuration. The prototype of D flip-flop with GNR edge-contact (Fig. S10) shows obvious practical advantages in energy efficiency compared with PCRAM cells made from silicon transistor. A narrow GNR not only brings the blade contact to PCM, but also serves as channel of a field effect transistor with high on/off ratio. Phase change devices with edge contact of GNRs have shown their potential as the building blocks of unconventional computing architectures to bypass the von Neumann bottleneck[32,33] and overcome the limits of Dennard scaling[34].

**Methods:**
**Device fabrication.**
Both $h$-BN and graphene flakes are prepared from their bulks by mechanical exfoliation[35]. A pick-up technique similar to the literature[36] is adapted to sandwich graphene in-between two $h$-BN layers. And a mask of PMMA resist is defined by lithography on the capping $h$-BN surface. The regions of hetero-structure outside of the mask are then etched away to expose the edge of graphene. Pd electrodes in a thickness of 50 nm are deposited by magnetron sputter. After another small window is opened to expose the edge of graphene, the GST material is deposited by magnetron sputter. A diagram of the preparation process is shown in Fig. S1.

The GNRs were grown by chemical vapor deposition[24]. The fabrication for D flip-flop based on the memory cell with edge-contact of GNR is shown in Fig. S10a. A flake of graphite is peeled onto the silicon substrate with 300 nm $SiO_2$. After that, $h$-BN, which is in suitable thickness (~50 to 60 nm), is then transferred onto the graphite. And then a $h$-BN flake with embedded GNRs is transferred onto the $h$-BN flake. After that, two metal leads are applied to GNRs while a metal lead is connected to graphite. Subsequently, a $h$-BN layer in a thickness of ~20 nm covers the GNR devices and acts as a protective layer. Finally, a window was opened by etching and then the GST with ~30 nm-thick is deposited into the window to connect the metal lead and GNR.

**GST deposition.**

GST films are deposited by using a sputtering system (ULVAC JEOL 7800F). The deposition rate is kept at ~0.6 Å/s at 20 W RF power to minimize the damage to the GNR or graphene.

**Electrical measurement**

Before electrical measurement, the memory devices were annealed at ~260 °C for 3 min in nitrogen flow using a RTP. Electrical measurements were performed with the combination of a Keithley 4200 Semiconductor Characterization System (SCS), an Arbitrary Waveform Generator (Tektronix AWG5002B), two Source Meters (Keithley 2400-C and 2502A). The device resistance is measured by applying a DC bias of 0.1 V. And the transient voltages were recorded with an oscilloscope (Tektronix MDO3032).

**Raman spectroscopy characterization**

The number of layer for graphene was determined by Raman spectroscopy with a excitation line of 532 nm, and the power of the laser is kept less than 1 mW.

**Atomic force microscopy (AFM)**

The thickness of graphite and *h*-BN were measured by an AFM (Dimension Icon, Bruker) in tapping mode. And the GNRs was located by AFM before encapsulating by capping layer of *h*-BN.

**TEM and STEM characterization**

The film of GST after annealing at 260 °C for 3 min was characterized by TEM (JEOL 2100F). And the cross sectional of edge-contact and the width of GNRs were performed in a double Cs-corrected TEM/STEM (JEM-ARM300F, JEOL) instrument operated at 80 kV.

**Acknowledgement**

H. Wang and X. Xie thank J.H. Edgar (Kansas State University, USA) for supplying partial of the *h*-BN crystals. X. Wang thanks Xiaoyu Liu and Hao Wang for assistance in device fabrication. X. Wang thank Shilong Lv and Tianjiao Xin (Microstructural Characterization Platform in Shanghai Institute of Microsystem and Information Technology, Chinese Academy of Sciences) for lamellae preparation and STEM measurement. We also thank F. Rao from Shenzhen Univ. for discussions.

**Funding:**

The work was partially supported by the National Key R&D program (Grant Nos. 2017YFF0206106, 2017YFA0206101), the Strategic Priority Research Program of Chinese Academy of Sciences (Grant No. XDB30000000), the National Natural Science Foundation of China (Grant Nos. 51772317, 91964102, 12004406, 91964204, 61874129), the Science and Technology Commission of Shanghai Municipality (Grant No. 18511110700, 20DZ2203600), Shanghai Rising-Star Program (A type) (Grant No.18QA1404800), the Shanghai Post-doctoral Excellence Program, the China

Postdoctoral Science Foundation (Grant Nos. 2017M621563, 2018T110415 2019T120366, 2019M651620, BX2021331), Shanghai Sailing Program (Grant No. 20YF1456400). Soft Matter Nanofab (SMN180827) of ShanghaiTech University. K.W. and T.T. acknowledge support from the Elemental Strategy Initiative conducted by the MEXT, Japan, Grant Number JPMXP0112101001, JSPS KAKENHI Grant Number JP20H00354 and the CREST(JPMJCR15F3), JST.

**Author Contributions:**
H. Wang conceived and designed the research. H. Wang, Z. Song and X. Xie supervised the research work. X. Wang fabricated the devices and carried out electrical measurements. L. Chen, H. S. Wang, C. Chen and C. Jiang prepared the GNRs, S. Song, T. Guo, Y. Xue and R. Wang performed the sputtering deposition of GST. C. Chen and C. Jiang performed AFM measurements. T. Guo and Y. Xue carried out the TEM measurements. S. Zhang and W. Song performed the modeling calculation of heat distribution for memory cell with graphene edge-contact. H. Wang, S. Song and X. Wang analyzed the experimental data and designed the figures. H. Wang, X. Wang and S. Song co-wrote the manuscript, and all authors contributed to critical discussions of the manuscript.